\documentstyle[12pt]{article}

\def\reff #1{(\ref{#1})}
\def\bra #1{\langle{#1}|}
\def\ket #1{|{#1}\rangle}
\def\phan{\vphantom{\frac12}}

\def\non{\nonumber}
\def\al{\alpha}
\def\be{\beta}
\def\ga{\gamma}
\def\la{\lambda}
\def\La{\Lambda}
\def\D{{\cal D}}
\def\L{{\cal L}}
\def\de{\delta}
\def\cP{{\cal P}}
\def\({\left(}
\def\){\right)}
\def\l{\left\{}
\def\r{\right\}}

\def\beqn{\begin{eqnarray}}
\def\eeqn{\end{eqnarray}}
\def\beq{\begin{equation}}
\def\eeq{\end{equation}}
\author{S. M. Klishevich%
 \thanks{E-mail address: klishevich@mx.ihep.su} \\
         {\it  Institute for High Energy Physics
         } \\
         {\it Protvino, Moscow Region, 142284, Russia
         }
       }
\title{Massive Fields with Arbitrary Integer Spin in Homogeneous
Electromagnetic Field}

\begin{document}
\maketitle

\begin{abstract}\noindent
We study the interaction of gauge fields of arbitrary integer spins with
the constant electromagnetic field. We reduce the problem of obtaining the
gauge-invariant Lagrangian of integer spin fields in the external field to
purely algebraic problem of finding a set of operators with certain
features using the representation of the high-spin fields in the form of
vectors in a pseudo-Hilbert space. We consider such a construction up
to the second order in the electromagnetic field strength and also present
an explicit form of interaction Lagrangian for a massive particle of spin
$s$ in terms of symmetrical tensor fields in linear approximation.
The result obtained does not depend on dimensionality of space-time.
\end{abstract}
\newpage
\section{Introduction}
\noindent
In spite of the long history \cite{Dirac-36,Fierz,Rarita} the problem of
obtaining the consistent interaction of high-spin fields is far from its
completion now.

The problem of obtaining the consistent description of "minimal"
interactions of high-spin fields with Abelian vector field plays a
particular role as well as gravitational interactions of such fields. In a
sense these interactions are the test ones since they allow one to connect
the fields of high spins with the observable world.

It has been realized \cite{Zinoviev-2} that one cannot build the
consistent "minimal" interaction with an Abelian vector field for the
massless fields of spins $s\ge\frac32$ in an asymptotically flat space-time.
The same is valid for the gravitational interaction of fields with spin
$s\ge2$ \cite{Aragone:79PL}. It is possible to argue the given statement as
follows \cite{Zinoviev-2}: The free gauge-invariant Lagrangian for integer
spin fields in the flat space has the structure
$\L_0=\partial\Phi\partial\Phi$
with transformations $\de\Phi=\partial\xi$. The introduction of the
"minimal" interaction means the replacement of the usual derivative with the
covariant one $\partial\to\D$. The gauge invariance fails and a
residual of the type $\left[\D,\D\right]\D\Phi\xi={\cal R}\D\Phi\xi$
appears, where ${\cal R}$ is the strength tensor of the electromagnetic
field or Riemann tensor. In the case of the electromagnetic interaction
for the fields with spins $s\ge\frac32$, one cannot cancel the residual by
any changes of the Lagrangian and the transformations in the linear
approximation. Therefore, in such a case this approximation does not exist,
but since linear approximation does not depend on the presence of any other
fields in the system. This means that the whole theory of interaction does
not exist either. The same is valid for the fermionic fields. In the case
of the gravitational interaction the residual for the field with spin
$\frac32$ is proportional to the gravity equations of motion:
$\de\L_0\sim i(\bar\psi^\mu\ga^\nu\eta)(R_{\mu\nu}-\frac12g_{\mu\nu}R)$. 
One can compensate such a residual by modifying the Lagrangian and the
transformations. As a result, the theory of the supergravity appears.
For fields with spins $s>2$ the residual contains terms proportional to the
Riemann tensor $R_{\mu\nu\,\al\be}$. It is impossible to cancel such terms
in an asymptotically flat space. Hence, the gravitational interaction does
not exist for any massless fields with spin $s>\frac32$. So, even if the
massless high-spin fields possess a nontrivial self-action, in any case
they are "thing in itself".

These difficulties can be overcome in several ways. In case of the
gravitational interaction, one can consider the fields in a constant
curvature space. Then, the Lagrangian for gravity would have an additional
term $\Delta\L\sim\sqrt{-g}\la$, where $\la$ is the cosmological constant. 
Modification of the Lagrangian and transformations leads to mixing of
terms with different numbers of derivatives. This allows one to compensate
the residual with terms proportional to $R_{\mu\nu\al\be}$. The complete
theory will be represented as series in the inverse value of cosmological
constant \cite{Vas-DS-1,Vas-DS-2}. This means the non-analyticity of theory
with respect to $\la$ at zero, i.e. impossibility of a smooth
transition to the flat space. Such a theory was considered in
Refs.\cite{Vas-DS-1,Vas-DS-2,Vasilev:IJMP95}.

It is also possible to avoid these difficulties if one considers
massive high-spin fields \cite{Sing-Hagen,mass_spin}.

In literature the electromagnetic \cite{FPT-92} and gravitational
\cite{P-93,CDP-94} interactions of arbitrary spin fields were considered at
the lowest order. Under consideration of the interactions, the authors
start from the free theory of the massive fields in the classical form
\cite{Sing-Hagen}. The "minimal" introduction of the interaction leads to
contradictions, therefore, it is necessary to consider non-minimal terms in
the interaction Lagrangian. Since the massive Lagrangian for spin-$s$ 
fields \cite{Sing-Hagen} is not gauge invariant, in such an approach there
are no restrictions of the form of non-minimal interaction and it is
necessary to introduce additional restrictions in order to build
a consistent theory. For instance, when studying the electromagnetic
interaction \cite{FPT-92}, the authors have used the requirement that
tree-level scatering amplitudes must possess a smooth $M\to0$ fixed-charge
limit in any theory describing the interaction of arbitrary-spin massive
particles with  photons. Under such requirement, the amplitudes do not
violate unitarity up to center-of-mass energies $E\gg M/e$. This restriction
leads to the gyromagnetic ratio $g=2$ for massive particles of any spin.
 When investigating the gravitational interactions \cite{P-93,CDP-94}, the
authors have required that tree-level amplitudes saturate the unitarity
bounds only at Planck scale.

It seems to us that it is more suitable to use the gauge-invariant approach
when one analyzes an interaction of massive fields
\cite{Zinoviev-83,Zinoviev-2,mass_spin,Pashnev-1,Argyres}. Under this
approach the interaction is considered as a deformation
\cite{Fronsdal-2,mass_spin} of initial gauge algebra and
Lagrangian\footnote{Of course, one must consider only a non-trivial
deformation of the free algebra and Lagrangian, which cannot be completely
gauged away or removed by a redefinition of the fields.}. Although,
generally speaking, the gauge invariance does not ensure the complete
consistency of massive theories but it allows one to narrow the search
and provides the appropriate number of physical degrees of freedom. Besides,
this approach is pretty convenient and practical.

At present only the superstring theory claims to have the consistent
description of interaction of the high-spin fields. But interacting
strings describe the infinite set of fields and the question about
interaction of finite number of fields is still open. In the case of the
constant Abelian field one can obtain a gauge-invariant Lagrangian that
describes the interaction for the fields of each string level. So, in
Refs.\cite{Argyres,spin3_1} the first and second massive levels of an open
boson string were investigated. These levels contain the massive fields of
spin 2 and 3. However, as was shown in Ref. \cite{spin3_1}, the presence of
the constant electromagnetic field leads to mixing of states on each string
level. Therefore, it is impossible to obtain an electromagnetic interaction
for a single field of spin $s$ in such approach.

In this paper we consider the interaction of an arbitrary massive field of
spin $s$ with homogeneous electromagnetic field up to the second order in
the strength.

We represent a free state with an arbitrary integer spin $s$ as state
$\ket{\Phi^s}$ in a Pseudo-Hilbert space\footnote{The representation of 
free fields with arbitrary integer spins in such a form was considered in
Refs.\cite{Pashnev-1,Pashnev:MPL97}.}. Tensor fields corresponding to
the particle with spin $s$ are coefficient functions in the state
$\ket{\Phi^s}$.  We introduce a set of operators in the considered Fock
space. We define the gauge transformations and necessary constraints for
the state $\ket{\Phi^s}$ by means of these operators. The gauge-invariant
Lagrangian has the form of expectation value of the Hermitian operator,
which consists of these operators, in the state~$\ket{\Phi^s}$. 
Using this construction, in section~2 we obtain the gauge-invariant
Lagrangian describing the particles with arbitrary integer spins in the
massless and massive case in terms of the coefficient functions.

 In the considered approach the gauge invariance is a consequence of
commutation relations of the introduced operators. Introduction of
interaction by replacing usual derivatives with covariant ones leads
to a change of algebraic features of the operators and as a consequence
to the loss of the gauge invariance. The problem of restoring the invariance
is reduced to an algebraic problem of finding such modified operators 
depending on the electromagnetic field strength, which satisfy the same
commutation relations as initial operators in the absence of external
field. It is possible to argue existence of such operators from
the consideration that the gauge transformation algebra in the free case 
and in the presence of an external field is the same (trivial)\footnote{But
transformations for the massive fields are not trivial.}. However, we
should note that in the massless case one cannot realize such a
construction. For the massive theory (section~3) we construct a
set of operators having the algebraic features of free ones up to the second
order in strength. Besides, we give an explicit form of the interaction
Lagrangian in terms of the tensor fields for a special case of the
constructed linear approximation. In the Appendix we dwell on the case of
massive spin-2 field.

\section{Free Field with Spin $s$}\label{Free}
\noindent
\subparagraph{Massless fields.}
 Let us consider the Fock space generated by creation and annihilation
operators $\bar a_\mu$ and $a_\mu$ which are vectors in the
{$D$-dimensional} Minkowski space ${\cal M}_D$ and which satisfy the
following algebra
\begin{eqnarray}
\label{H-alg}
\left[a_\mu,\bar a_\nu\right]&=&g_{\mu\nu}, \quad 
a_\mu^{\dag} = \bar a_\mu,
\end{eqnarray}
where $g_{\mu\nu}$ is the metric tensor with signature $\|g_{\mu\nu}\|={\rm
diag}(-1,1,1,...,1)$. Since the metric is indefinite, the Fock space, which
realizes the representation of the Heisenberg algebra \reff{H-alg}, is a
Pseudo-Hilbert space.
 
Let us consider the state in the introduceded space of the following type:
\begin{equation}
\label{Fstate}
\ket{\Phi^s}=\frac1{\sqrt{s}}\Phi_{\mu_1\dots\mu_s}(x)
\prod_{i=1}^s\bar a_{\mu_i}\ket{0}.
\end{equation}
Coefficient function $\Phi_{\mu_1\dots\mu_s}(x)$ is a symmetrical tensor
of rank $s$ in space ${\cal M}_D$. For this tensor field to describe the
state with spin\footnote{We consider symmetric tensor fields only.} $s$ 
one has to imposes the condition:
\begin{equation}
\label{2traceless}
\Phi_{\mu\mu\nu\nu\mu_4...\mu_s}=0.
\end{equation}
In terms of such fields Lagrangian \cite{Fronsdal-1,Curt1} has the form
\begin{eqnarray}
\label{LB-free}
\L_s&\!=\!&\frac12(\partial_\mu\Phi^s)
  \cdot(\partial_\mu\Phi^s)
  -\frac{s}{2}(\partial\cdot\Phi^s)\cdot(\partial\cdot\Phi^s)
  -\frac{s(s-1)}4(\partial_\mu\Phi'^s)
   \cdot(\partial_\mu \Phi'^s)  
\non\\&&{}\!\!\!\!
  -\frac{s(s-1)}{2}(\partial\cdot\partial\cdot\Phi^s)\cdot\Phi'^s
  -\frac18s(s-1)(s-2)(\partial\cdot\Phi'^s)
   \cdot(\partial\cdot\Phi'^s)
\end{eqnarray}
The following notation $\Phi'=\Phi_{\mu\mu...}$ is used here while the
point means the contraction of all indexes
$\Phi^s\cdot\Phi^s\stackrel{def}{=}
\Phi_{\mu_1\ldots\mu_s}\Phi^{\mu_1\ldots\mu_s}$.
This Lagrangian is invariant under the transformation
\begin{eqnarray}
\label{Gauge-0}
\de\Phi_{\mu_1\dots\mu_s}&=&\partial_{(\mu_1}\La_{\mu_2\dots\mu_{s-1})}, 
\\\label{tr-less}
\La_{\mu\mu\mu_3...\mu_{s-1}}&=&0.
\end{eqnarray}

Let us introduce the following operators in our pseudo-Hilbert space
\begin{equation}
\label{L-operator0}
L_1=p\cdot a,\quad L_{-1}=L_1^{\dag},\quad
L_2=\frac12a\cdot a,\quad L_{-2}=L_2^{\dag},\quad L_0=p^2.
\end{equation}
Here $p_\mu=i\partial_\mu$ is the momentum operator that acts in the space
of the coefficient functions. 

Operators of such type appear as constraints of a two-particle system
 under quanti\-za\-tion\footnote{It is also possible to regard operators
\reff{L-operator0} as a truncation of the Virasoro algebra.}
\cite{Barducci}. Operators \reff{L-operator0} satisfy the 
commutation relations:
\begin{equation}
\label{L-algebra0}
 \begin{array}{rclrcl}
 \left[L_1,L_{-2}\right]&=&L_{-1},
&\quad\left[L_1,L_2\right]&=&0,\\
 \left[L_2,L_{-2}\right]&=& N + \frac{D}{2},
&\quad\left[L_0,L_n\right]&=&0,\\
 \left[L_1,L_{-1}\right]&=&L_0,
&\quad\left[N,L_n\right]&=&{}-nL_n,\quad n=0,\pm 1,\pm 2.
 \end{array}
\end{equation}
Here $N=\bar a\cdot a$ is a level operator that defines the spin of states.
So, for instance, for the state \reff{Fstate} we have
$$
N\ket{\Phi^s} =s\ket{\Phi^s}.
$$

In terms of operators \reff{L-algebra0} condition \reff{2traceless} can be
written as
\begin{equation}
\label{Trace-L}
(L_2)^2\ket{\Phi^s}=0,
\end{equation}
while gauge transformations \reff{Gauge-0} have the form 
\begin{equation}
\label{Gauge-L}
\de\ket{\Phi^s}=L_{-1}\ket{\La^{s-1}}.
\end{equation}
Here, the gauge state
$$
\ket{\La^{s-1}}=\La_{\mu_1...\mu_{s-1}}\prod_{i=1}^{s-1}\bar a_{\mu_i}
\ket{0}
$$
satisfies the condition
\begin{equation}
\label{traceless}
L_2\ket{\La}=0.
\end{equation}
This condition is equivalent to \reff{tr-less} for the coefficient
functions.

Lagrangian \reff{LB-free} can be written as an expectation value of a
Hermitian operator in the state \reff{Fstate}
\begin{equation}
\label{Llagr}
\L_s=\bra{\Phi^s}\L(L)\ket{\Phi^s}, \quad \bra{\Phi^s}=\ket{\Phi^s}^{\dag},
\end{equation}
where
\begin{eqnarray}
\label{L-action}
\L(L)&=&L_0-L_{-1}L_1-2L_{-2}L_0L_2
-L_{-2}L_{-1}L_1L_2
\non\\&&{}
+\left\{L_{-2}L_1L_1 + h.c.\right\}.
\end{eqnarray}

Lagrangian \reff{Llagr} is invariant under transformations \reff{Gauge-L}
as a consequence of the relation
$$
\L(L) L_{-1}=(...)L_2.
$$

\subparagraph{Massive fields}

Let us consider massive states of arbitrary spin $s$ in a similar
manner. For this we have to extend our Fock space by means of introduction
of a scalar creation and annihilation operators $\bar b$ and $b$, which
satisfy the usual commutation relations
\begin{equation}
\label{H-alg+}
\left[b,\bar b\right]=1,\quad b^{\dag}=\bar b.
\end{equation}

Operators \reff{L-operator0} are modified as follows:
\begin{equation}
\label{L-m}
L_1=p\cdot a + mb ,\quad L_2=\frac12\(a\cdot a + b^2\),
\quad L_0=p^2+m^2.
\end{equation}
Here $m$ is an arbitrary parameter that has dimensionality of mass.
In non-interacting case one can consider such a transition as the
dimensional reduction ${\cal M}_{D+1}\to{{\cal M}_D\otimes S^1}$ with the
radius of sphere $R\sim1/m$ (refer also to\cite{Pashnev-1,Pashnev:MPL97}).

We shall describe the massive field of spin $s$ as the following vector
in the extended Fock space:
\begin{equation}
\label{FockM}
\ket{\Phi^s}=\sum\limits_{n=0}^s\frac1{\sqrt{n!(s-n)!}}
\Phi_{\mu_1\dots\mu_{n}}(x)\bar b^{s-n}\prod_{i=1}^n\bar a_{\mu_i}\ket{0}.
\end{equation}
The same as in the massless field case, this state satisfies condition
\reff{Trace-L} in terms of operators \reff{L-m}. The algebra of
operators \reff{L-algebra0} changes weakly, the only commutator modified is
\begin{equation}
\label{comL2m}
 \left[L_2,L_{-2}\right]=N + \frac{D+1}{2}.
\end{equation}
Here, as in the massless case, the operator $N=\bar a\cdot a + \bar bb$
defines the spin of massive states. The Lagrangian describing the massive
field of spin $s$ also has the form \reff{L-action}, where the
expectation value is taken in the state \reff{FockM}. Such Lagrangian is
invariant under transformations \reff{Gauge-L} with the gauge Fock vector
$$
\ket{\La^{s-1}}=\sum\limits_{n=0}^{s-1}\frac1{\sqrt{(n+1)!(s-n-1)!}}
\La_{\mu_1...\mu_n}\bar b^{s-n-1}\prod_{i=1}^n\bar a_{\mu_i}\ket{0},
$$
which satisfies condition \reff{tr-less}.

Having calculated expectation \reff{L-action} we obtain the explicit
expression for the Lagrangian describing the massive state with arbitrary
spin $s$ in terms of the coefficient functions
\begin{eqnarray}\label{L_s}
\L_0&=&
\sum_{n=0}^s\bar\Phi^n\cdot p^2\Phi^n\(1- C^2_{s-n}\)
-\sum_{n=2}^s\bar\Phi'^n\cdot p^2\Phi'^nC^2_n
\non\\&&{}
-\frac12\sum_{n=1}^s\(\bar\Phi^n\cdot p\)
\cdot\(p\cdot\Phi^n\)n\(2+C^2_{s-n}\)
-\frac32\sum_{n=3}^s\(\bar\Phi'^n\cdot p\)\cdot\(p\cdot\Phi'^n\)C^3_n
\non\\&&{}
-\l
\frac12\sum_{n=3}^s\(\bar\Phi'^n\cdot p\)\cdot\(p\cdot\Phi^{n-2}\)
\(n-2\)\sqrt{C^2_nC^2_{s-n+2}}\right.
\non\\&&{}
+\sum_{n=2}^s\bar\Phi'^n\cdot p^2\Phi^{n-2}\sqrt{C^2_nC^2_{s-n+2}}
-\sum_{n=2}^s\bar\Phi'^n\cdot\(p\cdot p\cdot\Phi^n\)C^2_n
\non\\&&\left.{}
-\sum_{n=2}^s\bar\Phi^{n-2}\cdot\(p\cdot p\cdot\Phi^n\)
\sqrt{C^2_nC^2_{s-n+2}}
 + h.c.\r
\non\\&&{}
-m\l
\frac12\sum_{n=1}^s\(\bar\Phi^n\cdot p\)\cdot\Phi^{n-1}
\(2-C^1_{s-n}+C^2_{s-n}\)\sqrt{nC^1_{s-n+1}} 
\right.\non\\&&{}
-\frac14\sum_{n=2}^s\bar\Phi'^n\cdot\(p\cdot\Phi^{n-1}\)
(n-1)(4-C^1_{s-n})\sqrt{nC^1_{s-n+1}}
\non\\&&{}
+\frac32\sum_{n=3}^s\(\bar\Phi'^n\cdot p\)\cdot\Phi^{n-3}
\sqrt{C^3_nC^3_{s-n+3}}
\non\\&&{}\left.
+\frac12\sum_{n=3}^s\(\bar\Phi'^n\cdot p\)\cdot\Phi'^{n-1}
C^2_n\sqrt{nC^1_{s-n+1}}
 + h.c.\r
\non\\&&{}
+\frac{m^2}2\biggl\{
\sum_{n=0}^s\bar\Phi^n\cdot\Phi^n
\(2-2C^1_{s-n}+2C^2_{s-n}-3C^3_{s-n}\)
\non\\&&{}
-2\sum_{n=2}^s\l\bar\Phi'^n\cdot\Phi^{n-2}C^1_{s-n}\sqrt{C^2_nC^2_{s-m+2}}
+h.c.\r
\non\\&&{}
-2\sum_{n=2}^s\bar\Phi'^n\cdot\Phi'^nC^2_n(2+C^1_{s-n})
\biggr\}.
\end{eqnarray}
Here $\Phi^n$ denotes a symmetrical tensor field of rank $n$ and we use the
following notation {$C^m_n=\frac{n(n-1)...(n-m+1)}{m!}$}. Condition
\reff{Trace-L} for the fields has the following form in this case
\begin{equation}
\label{2tr_Phi}
\sqrt{C^n_sC^2_n}C^2_{n-2}\Phi''^n
+2\sqrt{C^{n-2}_s}C^2_{n-2}C^2_{s-n+2}\Phi'^{n-2}
+ \sqrt{C^{n-4}_s}C^2_{s-n+4}C^2_{s-n+2}\Phi^{n-4}=0,
\end{equation}
where $n=4,...,s$. Correspondingly, the gauge transformations have the form
\begin{equation}\label{d_s}
\de_0 \Phi^n=\left\{p\La^{n-1}\right\}_{s.}
+m\sqrt{\frac{s-n}{n+1}}\La^n.
\end{equation}
Here $\{...\}_{s.}$ denotes symmetrization over all indexes. At the
same time condition \reff{traceless} can be written as
\begin{equation}
\label{tr_La}
C^2_n\La'^n + C^2_{s-n+1}\La^{n-2}=0,\qquad n=2,...,s-1.
\end{equation}

Obviously, the dimensional parameter $m$ has the sense of mass of the
state. For convenience, hereinafter we assume $m=1$.

One can derive the same result from \reff{LB-free} by means of the
dimensional reduction \cite{SivaKumar-PR85a,SivaKumar-PR85b} of massless
theory.

It is worth noting that the massive higher-spin fields represented by
Lagrangian \reff{L_s} are described by the first-class constraints
only\footnote{The Hamilton formulation for the gauge-invariant description
of the massive fields with spins 2 and 3 was considered in Ref.
\cite{hamilton}.} from point of view of the Hamilton formulation. As is
well-known the "minimal" coupling prescription breaks the right number of
physical degrees of freedom. In the gauge manner of description this is
represented as breaking the gauge invariance and, as a consequence, breaking
the algebra of the first-class constraints. But if we can restore the gauge
invariance by some deformation of the Lagrangian and the transformation,
hence we can restore the algebra of the constraints and, as a consequence,
the right number of physical degrees of freedom. A construction of such a
type is considered in the next section.

\section{Electromagnetic Interaction of Massive Spin $s$ Field}\label{int}
\noindent
In this section we consider the interaction of gauge massive fields of
arbitrary integer spin $s$ with constant electromagnetic field.

We introduce interaction by means of the "minimal coupling", i.e. we
replace usual momentum operators with $U(1)$-covariant ones {$p_\mu
\to\cP_\mu$}. The commutator of covariant momenta defines the
electromagnetic field strength
\begin{equation}
\left[\cP_\mu,\cP_\nu\right]=F_{\mu\nu}.
\end{equation}
For convenience we included the imaginary unit and coupling constant into
the definition of strength tensor.

In the definition of operators \reff{L-m}, we replace the usual momenta
with the covariant ones as well. As a result the operators cease to obey
algebra \reff{L-algebra0}. Therefore, Lagrangian \reff{L-action} loses the
invariance under transformations \reff{Gauge-L}.

To restore algebra \reff{L-algebra0}, \reff{comL2m} let us represent
operators \reff{L-m} as normal ordered functions of creation and
annihilation operators as well as of the electromagnetic field strength
i.e.
$$
L_i=L_i\(\bar a_\mu,\bar b,a_\mu,b,F_{\mu\nu}\).
$$
As a matter of fact, this means that these operators belong to extended
universal enveloping algebra of Heisenberg algebra \reff{H-alg},
\reff{H-alg+}. The particular form of operators $L_i$ is defined from
the condition of recovering of commutation relations \reff{L-algebra0} and
\reff{comL2m} by these operators. We should note that it is enough to
define
the form of operators $L_1$ and $L_2$, since one can take following
expressions:
\begin{equation}
\label{def}
L_0\stackrel{{\rm def}}{=}\left[L_1,L_{-1}\right],\quad
N\stackrel{{\rm def}}{=}\left[L_2,L_{-2}\right]-\frac{D+1}2.
\end{equation}
as definitions of operators $L_0$ and $N$.

Since we have turned to the extended universal enveloping algebra, the
arbitrariness in the definition of operators\footnote{Such an arbitrariness
has been also presented earlier as an internal automorphism
of the Heisenberg-Weil algebra defining the Fock space \cite{Turbiner:97}.
But exactly the transformations depending on $F_{\mu\nu}$ are important for
us.} $a$ and $b$ appears. Besides, in the right-hand side of \reff{H-alg}
and \reff{H-alg+}, we can admit the presence of the arbitrary operator
functions depending on $a$, $b$, and $F_{\mu\nu}$. Here, such a
modification of the operators must not lead to breaking the Jacobi
identity. Besides, these operators must restore the initial algebra in limit
$F_{\mu\nu}\to 0$. However, one can make sure that using the arbitrariness
in the definition of creation and annihilation operators, we can restore
algebra \reff{H-alg}, \reff{H-alg+} at least up to the second order in the
strength.

We shall search for operators $L_1$ and $L_2$ in the form of an expansion
in the strength tensor which is equivalent to that in the coupling constant.

Let us consider the linear approximation.

Operator $L_1$ should be no higher than linear in operator $\cP_\mu$, since
the presence of its higher number changes the type of gauge transformations
and the number of physical degrees of freedom. Therefore, at this order we
shall search for operators having the form
\begin{equation}
\label{L1anz}
L_1^{(1)}=\(\bar aFa\)h_0(\bar b,b)b + \(\cP Fa\)h_1(\bar b,b)
+ \(\bar aF\cP\)h_2(\bar b,b)b^2.
\end{equation}
At the same time operator $L_2$ cannot depend on the momentum operators at
all, since condition \reff{Trace-L} defines purely algebraic constraints
on the coefficient functions. Therefore, at this order we choose operator
$L_2$ in the following form:
\begin{equation}
\label{L2anz}
L_2^{(1)}= \(\bar aFa\)h_3(\bar b,b)b^2,
\end{equation}
Here $h_i(\bar b,b)$ are normal ordered operator functions of the type
$$
h_i(\bar b,b)=\sum\limits_{n=0}^\infty H_n^i \bar b^n b^n,
$$
where $H^i_n$ are arbitrary real coefficients. We consider only the
real coefficients since the operators with purely imaginary coefficients do
not give any contribution to the "minimal" interaction.

Let us define the particular form of functions $h_i$ from the condition 
recovering commutation relations \reff{L-algebra0} by operators
\reff{L1anz} and \reff{L2anz}. This algebra is entirely defined by
\reff{def} and by the following commutators
\begin{eqnarray}
\label{l2l1}
[L_2,L_1]&=&0,
\\
\label{l2l-1}
[L_2,L_{-1}]&=&L_1,
\\
\label{l0l1}
[L_0,L_1]&=&0.
\end{eqnarray}
Having calculated \reff{l2l1} and passing to the normal symbols of creation
and annihilation operators, we obtain a system of differential equations
for the normal symbols of operator functions $h_i$. For the normal symbols
of the operator functions we shall use the same notations. This does not
lead to the mess since we consider the operator functions as the functions
of two variables while their normal symbols as the functions of one
variable. From \reff{l2l1} we have the equations
\begin{eqnarray}
\label{Dif_eq1}
&&\frac12h_2''(x)+h_2'(x)=0,
\non\\
&&\frac12h_0''(x)+h_0'(x)-h_3'(x)=0,
\\\non
&&\frac12h_1''(x)+h_1'(x)-h_2(x)-h_3(x)=0.
\end{eqnarray}
Here the prime denotes the derivative with respect to $x$, while
$x=\bar\be\be$, where $\bar\be$ and $\be$ are the normal symbols of
operators $\bar b$ and $b$, correspondingly.

Similarly, from \reff{l2l-1} we derive another system of the equations:
\begin{eqnarray}
\label{Dif_eq2}
&&\non\phan
x^2\biggl(\frac12h_2''(x)+h_2'(x)\biggr)+2x\(h_2'(x)+h_2(x)\)
+h_2(x)-2h_1(x)=0,
\\&&
\frac12h_1''(x)+h_1'(x)-h_2(x)+h_3(x)=0,
\\&&\non\phan
x\biggl(\frac12h_0''(x)+h_0'(x)+h_3'(x)\biggr)+h_0'(x)+2h_3(x)=0.
\end{eqnarray}
Having solved the systems of equations \reff{Dif_eq1} and \reff{Dif_eq2} we
obtain the particular form of functions $h_1$:
\begin{eqnarray}
\label{resh_i}
h_0(x)&=&{\rm const},
\non\\
h_1(x)&=&d_1\(\frac12-x\)e^{-2x}+d_2\(\frac12+x\),
\non\\
h_2(x)&=&d_1e^{-2x}+d_2,
\\\non
h_3(x)&=&0.
\end{eqnarray}
Here $d_1$ and $d_2$ are arbitrary real parameters. Using \reff{resh_i}
we obtain from \reff{l0l1} 
$$
h_0(x)=1-d_2.
$$
The transition to the operator functions is realized in the conventional
manner:
$$\left.
h_i(\bar b,b)=\exp\(\bar b\frac{\partial}{\partial \bar \be}\)
\exp\(b\frac{\partial}{\partial \be}\) h_i(\bar\be\be)
\right|_{{\bar\be\to0}\atop{\be\to0}}=\ :\!h_i(\bar bb)\!:.
$$

So, normal symobls of operators $L_n$ has the following form in this
approximation:
\begin{eqnarray*}
\label{L^(1)}
L_1^{(1)}&=&
(1 - d_2)\(\bar\al F\al\)\be
+ \(e^{-2\bar\be\be}d_1\(\frac12 - \bar\be\be\)
 + d_2\(\frac12 + \bar\be\be\)\)(\cP F\al)
\\&&{}
 + \(e^{ - 2 \bar\be\be} d_1 + d_2\)\(\bar\al F\cP\)\be^2,
\\
L_0^{(1)}&=&
 (1 - 2 d_2)\(\bar\al F\al\)
 + \l (1 + 2 d_2)\(\cP F\al\)\bar\be + h.c.\r,
\\
L_2^{(1)}&=&0,
\end{eqnarray*}
where $\bar\al_\mu$ and $\al_\mu$ are the normal symbols of operators~$\bar
a_\mu$ and~$a_\mu$. Since operator $L_2$ has not changed in the linear
approximation, hence from \reff{def} it follows that operator $N$ and
constraints \reff{2tr_Phi}, \reff{tr_La} have not changed either.

 Thus, we have obtained the general form of operators $L_n$ which obey
algebra \reff{L-algebra0} in the linear approximation. This means that
Lagrangian \reff{L-action} is an invariant under
transformations~\reff{Gauge-L} at this order.

From \reff{L^(1)} it is clear that there exists the two-parametric
arbitrariness in the linear approximation. But one of the arbitrary
parameters $d_1$ and $d_2$ is determined in the second approximation
In this, there are two solutions: when $d_1$ vanishes and $d_2$ is
arbitrary, and vice versa, when $d_1$ is a free parameter and $d_2$ is
equal to $\frac12$. One can verify that the gyromagnetic ratio vanishes in
the second case. Below we will consider the first solution only. Having set
$d_1=0$ we calculate expectation value \reff{L-action} and obtain the
linear in $F_{\mu\nu}$ expression of the Lagrangian describing the e.m.
interaction of massive field of spin $s$ in terms of the coefficient
functions:
\begin{eqnarray*}
\label{Lagr1}
\L^{(1)}&\!\!=\!\!&
\sum_{n=1}^s\bar\Phi^n\!\cdot\! F\!\cdot\!\Phi^n
n\(1-2d_2+2\(d_2-1\)C^1_{s-n}+\(3-2d_2\)C^2_{s-n}
\non\right.\\&&\left.{}
+3\(d_2-1\)C^3_{s-n}\)
+3\sum_{n=3}^s\bar\Phi'^n\!\cdot\! F\!\cdot\!\Phi'^n
C^3_n\(2d_2-1+\(d_2-1\)C^1_{s-n}\)
\\&&{}
+\l
\sum_{n=3}^s\bar\Phi'^n\!\cdot\! F\!\cdot\!\Phi^{n-2}
\sqrt{C^2_nC^2_{s-n+2}}(n-2)\(1+\(d_2-1\)C^1_{s-n}\)
\non\right.\\&&{}
+ \frac12\sum_{n=2}^s\(\cP\!\cdot\!\bar\Phi^n\)\!\cdot\!
F\!\cdot\!\Phi^{n-1}\sqrt{nC^1_{s-n+1}}(n\!-\!1)(d_2\!-\!1)\!
\(2-2C^1_{s-n}\!+C^2_{s-n}\)
\non\\&&{}
+\!\sum_{n=1}^s\!\(\bar\Phi^n\!\cdot\! F\!\cdot\!\cP\)\!\cdot\!\Phi^{n-1}
\sqrt{nC^1_{s-n+1}}\!\(2+3d_2-3d_2C^1_{s-n}\!-\!\(2-{\frac72}d_2\)
\!C^2_{s-n}
\non\right.\\&&\left.{}
-6d_2C^3_{s-n}\)
+\frac32\sum_{n=4}^s\(\cP\!\cdot\!\bar\Phi'^n\)\!\cdot\!
F\!\cdot\!\Phi'^{n-1}\sqrt{nC^1_{s-n+1}}C^3_{n-1}(d_2-1)
\non\\&&{}
+\frac12\!\sum_{n=2}^s\!\bar\Phi'^n\!\cdot\!\(\cP\!\cdot\!
F\!\cdot\!\Phi^{n-1}\)\sqrt{nC^1_{s-n+1}}(n\!-\!1)\!\(d_2\!-\!
\(1\!+\!\frac{d_2}4\)\!C^1_{s-n}\!-8C^2_{s-n}\)
\non\\&&{}
-\frac12\sum_{n=3}^s\bar\Phi'^n\!\cdot\! F\!\cdot\!
\(\cP\!\cdot\!\Phi^{n-1}\)\sqrt{nC^1_{s-n+1}}C^2_{n-1}(d_2-1)\(4-C^1_{s-n}\)
\non\\&&{}
-\sum_{n=3}^s\(\bar\Phi'^n\!\cdot\! F\!\cdot\!\cP\)\!\cdot\!\Phi'^{n-1}
\sqrt{nC^1_{s-n+1}}C^2_{n-1}\(1+\frac94d_2-d_2C^1_{s-n}\)
\non\\&&{}
-3\sum_{n=3}^s\(\bar\Phi'^n\!\cdot\! F\!\cdot\!\cP\)\!\cdot\!\Phi^{n-3}
\sqrt{C^3_nC^3_{s-n+3}}\(1+\frac{d_2}4C^1_{s-n}\)
\non\\&&{}
+ \frac32\sum_{n=4}^s\bar\Phi^{n-3}\!\cdot\! F\!\cdot\!
\(\cP\!\cdot\!\Phi'^n\)\sqrt{C^3_nC^3_{s-n+3}}(n-3)(d_2-1)
\non\\&&{}
-\frac{d_2}4\sum_{n=1}^s\(\bar\Phi^n\!\cdot\!
F\!\cdot\!\cP\)\!\cdot\!\(\cP\cdot\Phi^n\)
n\(2+4C^1_{s-n} - 7C^2_{s-n} + 6C^3_{s-n}\)
\non\\&&{}
+d_2\sum_{n=2}^s\bar\Phi'^n\!\cdot\!\(\cP\!\cdot\!
F\!\cdot\!\(\cP\!\cdot\!\Phi^n\)\)
C^2_n\(1+2C^1_{s-n}-C^2_{s-n}\)
\non\\&&{}
-\frac34d_2\sum_{n=3}^s\(\bar\Phi'^n\!\cdot\! F\!\cdot\!\cP\)\!\cdot\!
\(\cP\!\cdot\!\Phi'^n\) C^3_n\(1+2C^1_{s-n}\)
\non\\&&{}
- d_2\sum_{n=2}^s\(\(\bar\Phi^n\!\cdot\!\cP\)\!\cdot\!F\!\cdot\!\cP\)
\!\cdot\!\Phi^{n-2}\sqrt{C^2_nC^2_{s-n+2}}\(1-2C^1_{s-n}+C^2_{s-n}\)
\non\\&&{}
-\frac{d_2}2\sum_{n=4}^s\(\(\bar\Phi'^n\!\cdot\!\cP\)\!\cdot\!
F\!\cdot\!\cP\) \!\cdot\!\Phi'^{n-2}C^2_{n-2}\sqrt{C^2_nC^2_{s-n+2}}
\non\\&&{}
+\frac{d_2}4\sum_{n=3}^s\(\bar\Phi'^n\!\cdot\! F\!\cdot\!\cP\)\!\cdot\!
\(\cP\!\cdot\!\Phi^{n-2}\)\(n-2\)\sqrt{C^2_nC^2_{s-n+2}}\(7-2C^1_{s-n}\)
\non\\&&{}
-\frac{d_2}4\sum_{n=3}^s\(\bar\Phi'^n\!\cdot\!\cP\)\!\cdot\!
\(\cP\!\cdot\! F\!\cdot\!\Phi^{n-2}\)
\(n-2\)\sqrt{C^2_nC^2_{s-n+2}}\(1+2C^1_{s-n}\)
\non\\\non&&{}\left.
-3d_2\sum_{n=4}^s\(\(\bar\Phi'^n\!\cdot\!\cP\)\!\cdot\!F\!\cdot\!
\cP\)\!\cdot\!\Phi^{n-4}\sqrt{C^4_nC^4_{s-n+4}}
 + h.c.\r.
\end{eqnarray*}
Correspondingly, in this approximation the gauge transformations have the
form:
\begin{eqnarray*}
\de_1\Phi^n&=&\frac{d_2}2\(1+2\(s-n\)\)\l\(F\cdot\cP\)\La^{n-1}\r_{s.}
\non\\&&{}
+2d_2\(n+1\)C^2_{s-n}\sqrt{\frac{C^2_{s-n}}{C^2_{n+2}}}
\(\cP\cdot F\cdot\La^{n+1}\)
\\\non&&{}
+\(1-d_2\)n\(s-n-1\)\sqrt{\frac{s-n}{n+1}}\l F\cdot\La^n\r_{s.}.
\end{eqnarray*}
It is worth noting that the construction obtained is free from pathologies.
Indeed, the gauge invariance ensures the appropriate number of physical
degree of freedom. In this, the model is causal in linear approximation
since by virtue of the antisymmetry and homogeneity of $F_{\mu\nu}$ the
characteristic determinant\footnote{The determinant is entirely determined
by the coefficients of the highest derivatives in equations of motion after
gauge fixing and resolving of all the constaraints \cite{Velo1:69}.}
for equations of motion of any massive state has the form $D(n)=(n^2)^p +
{\cal O}\(F^2\)$, where $n_\mu$ is a normal vector to the characteristic
surface and the integer constant $p$ depends on the spin of massive state.
The equations of motion will be causal (hyperbolic) if the solutions $n^0$
to $D(n)=0$ are real for any $\vec n$. In our case condition $D(n)=0$
corresponds to the ordinary light cone at this order.

Let us consider the quadratic approximation in the strength. The same as at
preceding order, if one takes the general ansatz for operators $L_1$, $L_2$
and requires the recovering of relations \reff{L-algebra0}, we will
obtain a system of inhomogeneous differential equations at the second
order. As was stated above, this system has the two solutions for parameters
$d_1$ and $d_2$ and we choose the solution when $d_1$ vanishes and $d_2$ is
arbitrary. According to our choice, operators $L_1$ and $L_2$ have
the following form in this approximation:
\begin{eqnarray*}
L_1^{(2)}&=&{}
 \!\!\(\bar\al F\cP\)\(\bar\al F\al\)\!\be^2\(\frac14 - d_2^2 - d_2\)
 - \(\bar\al FF\bar\al \)\!\be^3 \(\frac13 e^{-2\bar\be\be} c_2+\frac12 d_2
+ \frac18\)
\\&&{}
 + \(\bar\al FF\cP\) \be^2 \(\frac13 e^{-2\bar\be\be} c_2 + 2 c_1
 - \frac12 d_2^2 - \frac12 d_2 + \frac18\)
\\&&{}
 + \(\bar\al F\al \)^2 \be \(d_2^2 + \frac12 d_2 - \frac58\)
 + F^2 \be \( c_3 - \frac18 \(1 + 4d_2\) \bar\be\be\)
\\&&{}
 + \(\bar\al FF\al \) \be 
\(\(d_2 + \frac14\) \bar\be\be + d_2^2 + \frac32 d_2 - \frac38\)
\\&&{}
 - \(\al FF\al\) \bar\be 
\(\frac13 c_2 e^{-2\bar\be\be}\(1-\bar\be\be\)
 + \frac18\(1 + 4d_2\)\(1+\bar\be\be\) + 2 c_1\)
\\&&{}
 + \(\bar\al F\al \) \(\cP F\al\) \(\bar\be\be + \frac12\)
\(\frac14 - d_2\(d_2 + 1\)\)
\\&&{}
 - \(\cP FF\al\) \(c_4 + \frac18\(4 d_2^2 + 4d_2 - 1\) \bar\be\be\),
\\
L_2^{(2)}&=&
 - \(\al FF\al\) \(\frac16 c_2 e^{-2\bar\be\be}\(1 - 2 \bar\be\be\)
 + c_1 \(1+2 \bar\be\be\)\)
\\&&{}
  + \(2\(\bar\al FF\al \) - F^2\) \be^2
  \(c_1 + \frac16e^{-2\bar\be\be} c_2\).
\end{eqnarray*}
Here $c_1$, $c_2$, $c_3$, and $c_4$ are arbitrary real parameters.

In contrast to the preceding order, operator $L_2$ and, correspondingly,
$N$ and constraints \reff{2tr_Phi}, \reff{tr_La} depend on $F_{\mu\nu}$ in
this approximation.

Operator $L_0$ is defined via the commutator of operators $L_1$, $L_{-1}$.
The part that is proportional to $F^2$ has the form:
\begin{eqnarray*}
L_0^{(2)}&=&{}
  \(\cP FF\cP\) 
\(\frac14 d_2^2 - 2 c_4 + \(d_2^2 - d_2 + \frac14\) \bar\be\be\)
\\&&{}
 + \(\bar\al F\cP\)\(\cP F\al\)\(d_2^2 - d_2 + \frac14\)\(1+2\bar\be\be\)
\\&&{}
 + \(\bar\al F\al\)^2 \(3 d_2^2 - d_2 - \frac14\)
 + \(\bar\al FF\al\) \(\(2 d_2 + 1\) \bar\be\be + 3 d_2^2 + \frac14\)
\\&&{}
 - F^2 \(\frac12 \bar\be\be - 2 c_3 - \frac12 d_2\)
 + \biggl\{
 \(\bar\al F\cP\)^2 \be^2 \(d_2^2 - d_2 + \frac14\)
\\&&{}
 - \(\al FF\al\)\bar\be^2 \(d_2 + \frac12\)
 - \(\bar\al F\cP\) \(\bar\al F\al\) \be \(4 d_2^2 - d_2 + \frac12\)
\\&&{}
 - \(\bar\al FF\cP\) \be \(2 d_2^2 - \frac12 d_2 + \frac14\)
 + h.c.\biggr\}.
\end{eqnarray*}

Thereby, we have restored algebra \reff{L-algebra0} up to the second order
in the electromagnetic field strength. It means that we restored the gauge
invariance\footnote{Which ensures the right number of physical degree of
freeedom if we do not consider terms with higher number of derivatives.} of
Lagrangian \reff{L-action} at the same order as well. Here
we have not used the dimensionality of space-time anywhere explicitly, i.e.
the obtained expressions do not depend on it.

\section{Conclusion}
\noindent
In this paper we have constructed the Lagrangian describing the
interaction of massive fields of arbitrary integer spins with the
homogeneous electromagnetic field up to the second order in the strength.
It is noteworthy that unlike the string approach \cite{Argyres} our
consideration does not depend on the space-time dimensionality, and,
moreover, we have described the interaction of the single field with
spin $s$ while in the string approach the presence of constant
electromagnetic field leads to the mixing of states with different spins
\cite{spin3_1} and one cannot consider any states separately.

We should note that the case of a constant Abelian field can be easily
extended to a non-Abelian case in linear approximation. In this case we have
to consider the external field as the covariantly constant one. Here, one
should take the whole vacuum as $\ket{0}\otimes e^i$, where $e^i$ are a
basis vectors in space of linear representation of a non-Abelian group. The
covariant derivative has the form $\partial_\mu+A_\mu^aT^a$, where $T^a$ are
operators realizing the representation. In linear approximation such a
modification does not change the algebraic features of our scheme and,
therefore, all the derived results are valid in this case as well.

\section*{Acknowledgment}
\noindent
The author is grateful to Yu. M. Zinoviev for the numerous useful
discussions and his help with the work.

\appendix
\section{Massive spin-2 field.}
Here we consider the propagation of the massive spin-2 field in the constant
electromagnetic background. The following state in the Fock space
corresponds to such field
$$
\ket{2}=\{\(\bar a\cdot h\cdot \bar a\)
+\(v\cdot\bar a\)\bar b + \varphi b^2
\}\ket{0}.
$$
It is easy to see that this state trivially satisfies
condition\footnote{One can verify that condition \reff{Trace-L} imposes 
not-trivial restrictions only on states with spin $4$ and higher.}
\reff{Trace-L}.

Having calculated the expectation value of operator \reff{L-action} in
this state we derive the following Lagrangian in linear approximation
\begin{eqnarray}\label{LagrS2}
\L&=&{}
 -2\bar h_{kl}D^2h_{kl}+2\bar hD^2h+4\bar
h_{lm}D_{lk}^2h_{km}-2\left(\bar h_{kl}D_{kl}^2h
 + \bar h_{kl}D_{kl}^2\varphi
-\bar hD^2\varphi +h.c.\right)
\non\\&&{}
-\bar v_kD^2v_k+\bar v_kD_{kl}^2v_l
 +2i\(\bar v_kD_kh-\bar v_lD_kh_{kl} - h.c.\)
 +2(\bar h_{kl}h_{kl}-\bar hh) 
\non\\&&{}
 -d_2\biggl( 2F_{km}\bar hD_{lk}^2h_{lm}
+2F_{km}\bar h_{mn}D_{kl}^2h_{ln}
+2F_{km}\bar h_{lm}D_{kl}^2\varphi
 +\frac 32F_{km}\bar
v_mD_{kl}^2v_l+h.c.\biggr)  
\non\\&&{}
 +i\Bigl( (3d_2+2)F_{kl}\bar v_mD_kh_{lm}
+2(d_2-1)F_{lm}\bar
h_{kl}D_kv_m+(d_2+1)F_{kl}\bar hD_kv_l
\non\\&&{}
 + 3F_{kl}\bar \varphi D_kv_l)-h.c.\Bigr)
 +4(2d_2-1)F_{kl}\bar
h_{lm}h_{km}+F_{kl}\bar v_lv_k,
\end{eqnarray}
where $h=g^{kl}h_{kl}$.

The gauge vector for the massive spin-2 state is 
$$
\ket{\La,2}=\{\(\xi\cdot\bar a\) + \eta b\}\ket{0}.
$$
Condition \reff{traceless} gives non-trivial constraints for gauge
vectors of massive states with spin $3$ and higher only.

From \reff{Gauge-L} we obtain the following gauge transformations for
Lagrangian \reff{LagrS2} in this order
\begin{eqnarray*}
\delta h_{kl} & = & iD_{(k}\xi_{l)}+i
\frac{d_2}2F_{(k|m}D_m\xi_{|l)}, \\ \delta v_k & = & iD_k\eta +\xi_k+i\frac
32d_2F_{km}D_m\eta +(1-d_2)F_{km}\xi_m, \\ 
\delta \varphi  & = & \eta +id_2F_{kl}D_k\xi_l.
\end{eqnarray*}

Let us fix the gauge invariance by means of the gauge condition
$$
v_\mu =0,\qquad \varphi=0.
$$
Now we have the following equations of motion
\beqn\label{S2eq}
0\ \,=\ \,\frac{\de \L_g}{\de\bar h_{kl}}&=&
g_{kl}\left( D^2h-D_{mn}h_{mn}-h-d_2F_{mp}D_{mn}^2h_{np}\right)
-D_{(kl)}^2h+2D_{(k|m}h_{m|l)}
\non\\&&{}
-D^2h_{kl}
+h_{kl} 
+d_2F_{mp}D_{(k|m}h_{p|l)}+d_2F_{m(k}D_{l)m}^2h)
\non\\&&{}
-d_2F_{m(k|}D_{mn}h_{n|l)}
+2(2d_2-1)F_{m(k|}h_{m|l)},
\eeqn
where $\L_g$ is Lagrangian \reff{LagrS2} in the gauge under
consideration. From these equations of motion we obtain the constraints:
\beqn\label{S2contr}
\non
D_lh_{kl}+d_2\left( \frac 32F_{lm}D_mh_{kl}-2F_{kl}D_mh_{lm}\right)  & = & 
0, \\ 
h & = & 0.
\eeqn
Thereby, we have the appropriate number of the constraints and,
respectively, the appropriate number of physical degree of freedom at
this order.

Now we briefly consider the question of causality of the massive spin-2
state in linear approximation. Using relations \reff{S2contr} and equations
of motion \reff{S2eq}, one can infer that the characteristic determinant has
the following form
$$
D(n)=\(n^2\)^{\frac{D^2(D+1)^2}{4}} + {\cal O}\(F_{kl}^2\).
$$
Thereby, from the condition $D(n)=0$, we obtain the usual light cone in this
approximation, i.e. the massive spin-2 state has the causal propagation in
the background under consideration.



\end{document}